\documentclass[english,a4paper,12pt]{article}
\usepackage{amssymb,graphicx,hyperref}

\begin{document}

\author{V.N. Lukash and V.N. Strokov\thanks{Astrospace Centre of the Levedev Physical Institute. E-mail: strokov@asc.rssi.ru}}
\title{Geometries with integrable singularity -- \\ black/white holes and astrogenic universes}
\date{}
\maketitle

\begin{abstract}
We briefly review the problem of generating cosmological flows of
matter in GR (the genesis of universes), analyze models'
shortcomings and their basic assumptions yet to be justified in
physical cosmology. We propose a paradigm of cosmogenesis based on
the class of spherically symmetric solutions with {\it integrable}
singularity $r=0$. They allow for geodesically complete geometries
of black/white holes, which may comprise space-time regions with
properties of cosmological flows.
\end{abstract}

\section{Introduction}\label{introduction}

I.D. Novikov's paper~\cite{Nov66} where he developed the model of
collapsing electrically charged sphere that gives rise to
expansion in another universe pioneered scientific research of the
cosmogenesis problem, that is, the problem of generating expanding
matter flows in General Relativity (GR). The studies took various
trends. Widely discussed became bouncing models and the birth of
the universe from ``nothing'' with matter entering a highly dense
state with de Sitter symmetry and then leaving this state by
tunnelling into the stage of cosmological expansion (see, for
example, \cite{Farhi87}-\cite{GMOV00}). The models with continuous
phase transitions turning collapse into expansion required matter
with an exotic equation-of-state, including the matter violating
the weak energy condition~\cite{Rub}. These and some other models
of cosmogenesis studied so far (see, for example, \cite{linde})
either started from the initial highly dense state with symmetry
close to that of the de Sitter space-time or assumed that the
similar state would emerge as a result of non-linear
quantum-gravitational effects, yet unknown to the modern science,
in the matter of collapsing object.

In this paper we propose another approach to the problem of
generating cosmological flows based on the class of general
spherically symmetric GR solutions with 2+2 split and {\it
integrable} singularity $r=0$. The latter allows us to continue
the radial geodesics with respect to the proper affine parameter,
thus, constructing geodesically complete geometries of black/white
holes. These solutions require the presence of effective matter in
$T$-regions {\it outside} the body of collapsing object and make
it clear, thanks to the continuous geodesics, that after the
inversion moment $r=0$ there forms an expanding white-hole
$T$-region with all the properties of cosmological flow. We show
that the eternal black/white hole metrics is sourced by an
effective matter with negative longitudinal pressure $p<0$ located
in the vicinity of the spatial hypersurfaces $r=0$. We present a
toy model of the geometry invariant with respect to the sign of
$r$ and make a suggestion that the effective matter with similar
properties also fills $T$-regions of astrophysical black holes.
This leads us to the following hypothesis of cosmogenesis: an
object collapsing into black hole generates ultra-high curvature
outside the collapsing body, which, in its turn, induces violent
particle creation, thus, transferring the initial momentum to the
newly created matter. The latter having passed to the $T$-region
of the white hole gives rise to an expanding cosmological flow. If
driven by inflation, the flow can grow to become quasi-Hubble in a
large volume. Such {\it astrogenic universes} may emerge inside
black/white holes which are naturally generated on the final stage
of evolution of stars, clusters and other compact astrophysical
objects in the maternal universe.

The next section deals with basic postulates and assumptions of
the cosmogenesis models. In Sects. 3 and 4 we study properties of
spherical solutions with respect to inversion of the spherical
reference frame and define classes of geodesically complete
geometries. Sects. 5 and 6 deal with the problem of the
black/white-hole source: we argue that the effective matter is
inevitably present in $T$-regions of the holes and develop toy
models of integrable singularities by introducing triggered phase
transitions. In the last section we summarize and discuss the
results.

\section{A brief review of the problem}\label{review}

Since GR came into being, the scientific community has had various
opportunities to verify that it gives a viable description of
phenomena that include strong gravitational fields and
relativistic velocities. Its experimental basis once consisting of
the three classical GR effects (perihelion precession, deflection
of light and red shift) has recently acquired one of its crucial
contributions -- the Cosmological Standard Model (CSM) of the
visible Universe. Although some theoretical premises of the model
(e.g. hypotheses of dark matter and dark energy) may inspire
certain GR modifications, in the conditions encountered in
observational cosmology Einstein's theory of gravity provides a
high degree of precision as soon as the CSM is challenged by
experimental data \cite{WMAP}.

The cornerstone of the standard model is the cosmological
principle synonymous to the large-scale Friedmannian symmetry.
Extrapolating the CSM to the past respects this symmetry and leads
to the geometry of the early Universe in the form of a rapidly
expanding quasi-Hubble matter flow that comes out of the state
with ultra-high space-time curvature and density \cite{lm}. If we
try to extrapolate this state within the general-relativistic
framework even closer to the ultimate zeroth instant of time, we
run into the singular point where the metric space itself does not
exist. The same happens in the black-hole physics where, however,
the singularity is, vice versa, the final stage of a collapsing
object.

As a rule, if a theory exhibits singularities they are a mere
consequence of neglecting some physical phenomena, or, in other
words, a consequence of idealization. Therefore, the singular
states in GR solutions indicate the limits of the modern theory of
gravity. The singularity problem arising in the GR (both in the
black-hole physics and cosmology) is caused by our inability to
fully understand what is gravitation and how it interacts with
matter. The theorems known from the 60's imply that world lines of
test particles cannot go beyond the singularity because of
infinite tidal forces \cite{hawkingpenrose}. The theorems are
based on some matter properties that are required {\it a priori}.
These are the dominant, strong and weak energy conditions. The
first two of them may not be fulfilled in quantum-gravity physical
processes where gravitation and matter intensely interact.
Besides,  one should differ the problem of existence of continuous
metric space-time itself (where a photon or test particle
propagate) and the problem of divergence of some curvature or
energy--momentum components. The current analysis demonstrates
that there are neither observational nor theoretical grounds
proving singular states inevitable.

Inflationary models \cite{guth} explained the large size and the
Friedmann symmetry of the observed matter flow, thus, providing a
theoretical basis for the cosmological principle itself. The
problem of initial conditions, however, continued unsolved. As a
matter of fact, one postulate (of isotropy and homogeneity) was
substituted by another. In other words, {\it where do the
ultra-high density and initial impetus launching the expansion
come from?} For example, in different models of
inflation~\cite{linde} new physical fields (often unknown to the
modern particle physics) are introduced in the ultra-dense state
from the very beginning. The birth of the Universe from "nothing"
also involves the notion of highly dense "false" vacuum. In
so-called bouncing models having been developed for more than 40
years now the problem of initial conditions is replaced for
specific matter properties at high densities, and again the
Friedmann symmetry is assumed.

In this paper we make use of the fundamental scientific principle
that states that any physical solution describing nature must
contain only such observable quantities that remain finite. It
leads us to consider realistic models of black/white holes with
{\it smoothed} metric singularities (the metric potentials remain
finite). This allows us to constrain the tidal forces (despite a
possible divergence of some curvature components) and construct a
geodesically complete metric space--time in dynamical models
incorporating solely the general physical principles --
energy--momentum conservation, a wide choice of the
equation-of-state, the weak energy condition. Thus, now the
geodesics pass to the $T$-region of a white hole rather than end
in the singularity. From this point we arrive to the hypothesis
that any black hole originated from the collapse of an
astrophysical object may give birth to a new (daughter or
astrogenic) universe. The conjecture naturally solves all of the
three above-mentioned problems of cosmogenesis:

\begin{itemize}
\item{The ultra-high curvature and density on the initial stage of
cosmological evolution are achieved as a consequence of
superstrong and highly variable gravitational fields that exist
inside the black/white hole and generate the proper matter of the
daughter universe.} \item{The initial push to the expansion of the
generated matter (the Big Bang) is given by the $T$-region of the
white hole. The initial cosmological impetus is, hence, of pure
gravitational nature and is one of the manifestations of
gravitational (tidal) instability.} \item{The $T$-region symmetry
of the black hole outside the maternal body of the collapsing
object is that of an anisotropic cosmology. It is transferred to
the white-hole $T$-region and can be made isotropic by known
inflationary mechanisms.}
\end{itemize}

None of unknown physical effects which are in play at high
densities can prevent us from using dynamical equations in the
form of the Einstein equations $G^{\mu}_{\nu}=8\pi
GT^{\mu}_{\nu}\,$, where any high-energy and high-curvature
geometrical modification is included in the right-hand side and
ascribed to the {\it effective} energy--momentum tensor
$T^{\mu}_{\nu}$, thus, containing both material and, in part,
spacetime degrees of freedom (see, for example,
\cite{sahni-starobinskii}). We do not calculate one-loop and other
"corrections" that are of no use in this case, since the processes
in question involve {\it non-linear} phase transitions taking
place in the effective matter at high curvature. We simulate this
non-linearity phenomenologically using dynamical solutions with
triggered matter generation and do not constrain the
equation-of-state. The basic geometrical object is the averaged
metric tensor $g_{\mu\nu}$\,, which is used to construct the
left-hand side of the equations according to the GR rules. The
equations obtained hold in the entire space--time, and the
manifold does not contain punctured points.

\section{Properties of geometries with respect to the inversion $r\rightarrow -r$}\label{spherical}

In the Schwarzschild metrics ($M$ is the external mass)
\begin{equation}
\label{schwarz}
ds^{2}=\left(1-\frac{2GM}{r}\right)dt^{2}-\frac{dr^{2}}{1-\displaystyle\frac{2GM}{r}}-r^{2}d\Omega\,,
\end{equation}
the quantity~$r>0$ \cite{frolov-novikov} plays two roles. On one
hand, it is the curvature radius of a 2-dimensional sphere with
the squared line element $r^{2}d\Omega$. On the other hand, it is
one of the coordinates. In order to discriminate one from the
other, let us present the metrics of an arbitrary spherically
symmetric 4D space--time split into a pair of 2D
spaces~\cite{polischik-2, GHP, lsm}:
\begin{equation}
\label{tau} dX^{\,2}= n_{IJ}\,dx^I dx^J
\end{equation}
and
\begin{equation}
\label{ell} dY^{\,2}=\gamma_{ij}\,dy^i dy^j \equiv r^2d\Omega\,,
\end{equation}
where the functions $n_{IJ}$ and $r$ depend on the variables
$x^I\!=(x^1,x^2)\in \mathbb{R}^2$ and are {\it independent} of the
internal 2D coordinates $y^i$ of the closed homogeneous and
isotropic 2-surface $\mathbb{S}^2$ of the unit curvature
$d\Omega\!= \omega_{ij}dy^i dy^j$, $\gamma_{ij}\!\equiv
r^2\omega_{ij}$. If the coordinates are chosen to be angular,
$y^i\!=(\theta,\varphi)$, we have $\omega_{ij}\!=
diag{\,(1,\,\sin^2\theta)}\,$ with $\,\theta\in [0, \pi]$ and
$\varphi\in [0, 2\pi)$.

By choosing the four coordinates of the covering grid one can turn
four non-diagonal components of the full metric tensor into zero
$g_{Ii}\!=0$ and write it in the orthogonal reference frame
$x^\mu\!=(x^I,y^i)$:
\begin{equation}
\label{s} ds^{\,2}= g_{\mu\nu}\,dx^\mu dx^\nu = dX^{\,2}-dY^{\,2},
\end{equation}
where $g_{\mu\nu}= diag{\,(n_{IJ}, -\gamma_{ij})}$ is the metrics
of the spherically symmetric geometry in the orthogonal split
$2+2$\,. The energy--momentum tensor corresponding to (\ref{s}) is
$T_{\mu\nu}\!=diag{\,(T_{IJ},-p_\perp\gamma_{ij})}\,$, \,where
$p_\perp$ is the transversal pressure.

At this point the metric potential $r$ in (\ref{ell}) can be
introduced in the invariant manner as a radius of the {\it
internal} curvature $\rho$ of the closed $Y$-space, where
\begin{equation}
\label{R2} R_{ij}^{(Y)}=\rho\gamma_{ij}\quad\; \mbox{and} \;\quad
\rho\equiv\frac 12 R^{(Y)}\!= r^{-2}
\end{equation}
are the Ricci tensor and scalar constructed from the metrics
$\gamma_{ij}$. By definition, the 2-space $\gamma_{ij}$ and its
internal curvature $\rho$ are invariant with respect to
interchanging $r$ and $-r$ while the {\it external} curvature of
$Y$ depends on the sign of $r$ and determines the orientation and
evolution of the surface in the space--time (\ref{s}):
\begin{equation}
\label{K} \mathcal{K}_{ijI}\equiv\frac 12 \gamma_{ij,I}=
\mathcal{K}_I \gamma_{ij}\,,\quad\mathcal{K}_I\equiv\frac 12
\gamma^{ij}\mathcal{K}_{ijI} =\frac{r_{,I}}{r}\,,
\end{equation}
where the comma in the subscript stands for the partial derivative
with respect to $x^I$. This fact becomes obvious in the
coordinates where one of the variables~$x^I$ is identically equal
to $r$.

Changing the sign $r\rightarrow -r$ with $\theta$ and $\varphi$
staying unchanged is equivalent to inverting\,\footnote{One can
fulfil the operation holding the condition $r\ge 0$, but
sacrificing the continuity of the angular coordinates at $r=0$:\,
$\theta\rightarrow\pi-\theta,\, \phi\rightarrow\phi+ \pi$. Note
that in the regions where $r$ does not change its sign one can
always restore the condition $r>0$ by renaming the coordinates
$\theta$ and $\varphi$.} the spherical reference frame (\ref{s}).
In the regions with negative $r$ with no matter ($T_{\mu\nu}=0$)
$r$ in the solution (\ref{schwarz}) is to be substituted with the
absolute value $|r|$. The parameter $M$ continues to stand for the
external (measured in the asymptotically flat space) hole mass.
The inversion results in the black hole transforming into the
white hole (and vice versa).

In the general case when the matter is present, the metrics
(\ref{s}) can be written in the Euler orthogonal coordinates:
\begin{equation}
\label{general-metrics}
ds^{2}=N^2(1+2\Phi)\,dt^{\,2}-\frac{dr^{\,2}}{1+2\Phi}-
r^2d\Omega\,,
\end{equation}
where $\Phi$ and $N$ are real finite functions of $(r,t)$. The
latter condition allows us to consider the metric fields on the
entire manifold $(r,t)\in\mathbb{R}^2$ without punctured points
and, thus, construct geodesically complete maps of the geometries
for {\it all} radial trajectories of test particles and light,
which are continued with respect to their proper affine
parameters. Recall that the solutions (\ref{general-metrics}) does
not diverge on the horizons $\Phi=-1/2$. It becomes obvious in the
Lagrange coordinates \cite{frolov-novikov}. By definition, in the
$T$($R$)-regions the potential $\Phi\,$ is less (greater) than
$-1/2$.

\section{Black/white holes and cosmology}\label{reversibility}

The different properties of the internal and external curvatures
with respect to inversion naturally classify the models. In the
$T$-regions, where $r$ is the time \cite{novikov}, (non)invariance
of (\ref{s}) with respect to the change of the $r$ sign is
(non)invariance of the metrics with respect to the time inversion.
The first class is that of reversible models while the second is
of {irreversible} ones.

The first type includes eternal black/white holes. Behavior of
geodesics in the space--time of the white hole is that of
geodesics in the black hole space--time corrected for the time
inversion (see Fig. 1). Every particle inside the black hole
horizon moves in the direction of increasing curvature while in
the white hole the curvature decreases on the
trajectory~\cite{frolov-novikov}. In particular, non-linear
quantum processes triggered by increasing intensity of the
gravitational field are reversible in this class of models. For
example, after being polarized vacuum experiences the same states
again in reverse order and returns to the initial state.
Identifying the gravitational field intensity with a
thermodynamical potential we can talk about an analogy with
reversible thermodynamical transitions.

Models of the second type deal with real irreversible processes
analogous to the irreversible phase transitions (say, to the
condensation of overheated vapor). Here the quantum-gravity
processes generating matter in the highly variable gravitational
field come into play. The matter does not disappear and remains in
the $T$-region \cite{zns,lm}. We show
(Sects.~\ref{model},~\ref{mode}) that in a certain class of models
the singularity emerging outside the collapsing matter gives birth
to a new world of the white hole which expands from high to low
densities (see Fig. 2). In the course of evolution phase
transitions occur and in some conditions inflation can be
realized. Driven by the latter, the cosmological expansion can
become isotropic within a large spatial volume. Therefore, among
irreversible models there are cosmological solutions, in which a
black hole (even with a small mass) gives birth to the
quasi-Hubble flow of an {\it astrogenic} universe.

It is worth pointing out that models of both the first and second
types incorporate significantly non-linear quantum processes which
cannot be treated with techniques of quasi-classical perturbation
theories of gravity. However, although the complete quantum theory
is yet to be developed, this cannot prevent us from simulating
this kind of processes phenomenologically on the basis of general
physical principles (see Sects.~\ref{introduction},~\ref{review}).
Here an analogy with classical hydrodynamics jumps to mind. Many
paradoxes of ideal fluid hydrodynamics were long
known~\cite{landafshits-6} to be solved by introducing the
coefficient of viscosity even though a microscopic viscosity
theory had not been developed yet.

\section{A material source of the Schwarzschild metrics}\label{model}

In the extended Schwarzschild metrics (so-called eternal
black/white hole) the singularity of black hole cannot source the
gravitational field in the outer space of the black hole, since it
lies in the absolute future relative to this space. Because of the
non-linearity of the relativistic equations of gravity the
following question is non-trivial -- {\it does GR require, as the
Newton gravity does, a central source of the curved vacuum
metrics, with its properties being found from the right-hand side
of the Einstein equations}\,\footnote{Recall that in the Newtonian
physics the spherically symmetric gravitational field in vacuum is
of the form $\Phi=-GM/r$, where $r= |\mathbf x|$. Evaluating the
right-hand side of the Poisson equation in the Euclidian space
$\mathbb{R}^3$, one finds that the source of this field is the
central mass $M=const$ located at $r=0$:\, $\Delta\Phi=4\pi
GM\delta^{(3)}(\mathbf x)$, where $\delta^{(3)}(\mathbf x)$ -- 3D
delta-function.}?

This problem has been discussed in the
literature~\cite{landafshits, MTW-3, Wald, frolov-novikov}, but an
agreement is yet to be reached and the problem remains unsolved.
Particularly, the authors ~\cite{balasin-nachbagauer} exploited a
formal mathematical approach applying the technique of integrating
tensor differential $n$-forms to evaluate the right-hand side.
After some type of regularization this results in appearing of
singular functionals (3D delta-functions), which, in turn, require
a space to live on. In~\cite{balasin-nachbagauer} this space is
identified with the locally defined Minkowski space of the
vierbein formalism. This choice is hard for us to agree with.
First, the required flat manifold must be defined globally and,
second, the singularity $r=0$ makes impossible even the local
definition because of the divergence of the vierbein.

Obviously, the issue of the extended Schwarzschild metrics source
in GR arises because of the singular spacelike hypersurfaces $r=0$
that completely reside in $T$-regions and are inherent to any
black/white hole. For this reason we cannot be satisfied with
"emergency solutions" which either remove the $T$-regions from
consideration at all (e.g. Einstein--Rosen
bridges~\cite{Poplawski}, wormholes~\cite{BrSt}) or modify them by
applying {\it specific constraints} on gravitation or matter
properties. Among the modifications are the requirements of finite
maximal curvature \cite{markov, markovfrolov} or matter density
\cite{dymnikova}, gravitational torsion \cite{Poplawski-1} and
others. I.G. Dymnikova constructed non-singular solutions with the
black/white hole asymptotics at large $r$ containing anisotropic
matter, which is 'vacuum-like' in the longitudinal direction and
constrained by the finite density condition (see~\cite{dymnikova}
and references therein). These solutions have at least {\it two}
$R$-regions, an external and internal one (the latter containing
the center $r=0$), separated by a $T$-region. Hence, their
topology dramatically differs from that of the Schwarzschild black
hole. These types of solutions are beyond the scope of our
consideration, since they lack the limit of the black/white hole
$T$-region.

Let us find out which properties of the effective matter support
the metric space--time. Making use of eq. (\ref{general-metrics})
one obtains from the GR equations \cite{landafshits,lsm}:
\begin{equation}
\label{rm} \Phi = -\frac{G m}{r}\,,
\end{equation}
where the finite {\it mass function}
\begin{equation}
\label{m} m=m\!\left(r,t\right)= 4\pi\!\int_{0} T_t^t r^2 dr= m_0
-4\pi r^2\!\!\int\! T_t^r dt
\end{equation}
vanishes on the inversion line $r=0$ (m(0,t) = 0)\, thanks to the
finiteness condition applied on the potential $\Phi$. The function
$m_0=m_0(r)$ is determined by the setting of the problem.
Therefore, GR informs us that the external mass in the
Schwarzschild solution exists {\it only if there is a material
source} of the metrics~(\ref{schwarz}). We also conclude that for
the function $m(r,t)$ to be finite, $T_t^t r^2$ must be integrable
at $r=0$. Hereafter, by definition, the distribution of pressure
over~$r$ integrable in this sense is called to possess an {\it
integrable} singularity. Mathematically, it is reminiscent of the
central cusp caustics in the spatial halo distribution of density
in the $R$-region.

Let the matter reside in a part of the $T$-region of the black
hole\,\footnote{Otherwise we would obtain a star or a halo.} with
curvature exceeding a critical value ($T_{\mu\nu}\ne 0$ when $0\le
r\le r_0<2GM$). Assume that at low curvature the space--time is
empty and described by the metrics (\ref{schwarz}) ($T_{\mu\nu}=0$
when $r > r_0$). This situation can be thought of as a {\it
triggered phase transition}, in which non-zero $T_\mu^\nu$
components emerge at a certain value of a curvature invariant that
increases near the singular hypersurface\,\footnote{One of the
candidates to play the role of the invariant is the squared
Riemann tensor
$\mathcal{I}=R_{\alpha\beta\gamma\delta}R^{\alpha\beta\gamma\delta}$
which in the Schwarzschild solution equals to $\,48(GM/r^3)^2$. In
this case the time moment $r_{0}$ when the phase transition starts
can be estimated from the physical dimension: $\mathcal{I}\sim
1/l_{P}^{4}$, where $l_{P}\approx 10^{-33}$~cm is the Planck
length. For a black hole with mass of the order of that of the Sun
$r_{0}\sim 10^{12}l_{P}$.}.

The effective matter distribution induced by the 4-curvature of
(\ref{general-metrics}) outside the body of the collapsing object
(hereafter, the 'star') respects the symmetry of the vacuum
gravitational field (\ref{schwarz}) and, thus, possesses the
global Killing $t$-vector. The matter is free from radial motion
($T_t^r=0$) and its distribution is homogeneous in the space
$(t,\theta,\phi)$ and extends to the white-hole $T$-region ($r<0$)
lying in absolute future with respect to the maternal black hole.
The generated matter in the white hole will remain in this
strictly homogeneous state till it interacts with the flow of the
particles coming from the surface of the collapsing 'star' (see
Fig. 2). In the homogeneity region the functions $\Phi, N,\,
p_\perp$, $\,T_I^J\!= diag\,(-p,\,\epsilon)\,$ and~$m$ depend
merely on time $r$ and we readily obtain a cosmological model with
the cylindrical symmetry $\mathbb{R}\times\mathbb{S}^{2}$ (the
Kantowski-Sachs model) capable of further dynamical isotropizing
to the Friedmann symmetry (see, for example,
\cite{lukash-starobinskii}). For this matter distribution the mass
function
\begin{equation}
\label{m0} m(r)=-4\pi\!\int_0 p\,r^2 dr
\end{equation}
is odd (asymmetric) with respect to  $r$ in models of the 1st
(2nd) type. The relation between $p\,$ and the black or white hole
mass\,\footnote{We assume that the matter in the white hole is in
the region $0< -r\le r_0\le 2GM$, where
$M=M_{WH}=-4\pi\!\int_0^{r_0}\!p(-r)\,r^2 dr$. In the 1st (2nd)
type models the black and white hole masses $M$ as well as the
$r_0$ parameters are equal (different).}
\begin{equation}
\label{m1} M=-4\pi\!\int_0^{r_0}\!p\,r^2 dr
\end{equation}
yields that if $M>0$, at the average the longitudinal pressure is
negative, $p< 0$. Note that eq. (\ref{m1}) does not constrain the
cosmological density $\epsilon\ge 0$. The weak energy condition
$\epsilon+p\ge 0$ yields the {\it lower} limit on the full mass
(per unit $t$ length) stored in either of the holes:
\begin{equation}
\label{m2} 4\pi\!\int_0^{2GM}\!\epsilon\,r^2 dr\ge M\,.
\end{equation}
A large universe in the white-hole $T$-region means both increased
age and mass. One can achieve it requiring\, $\Phi'>0$ ($Gm^\prime
>-\Phi > 1/2$, $-8\pi Gpr^2> 1$), which implies the inflation
condition.

The second metric potential $N$ and the transversal pressure
$p_{\perp}$ of the homogeneous matter are found from the rest of
the Einstein equations (the prime stands for the derivative with
respect to $r$):
\begin{equation}
\label{N} \frac{N^\prime}{N} =\frac{4\pi
Gr^2\!\left(\epsilon+p\right)}{2Gm-r}\,, \,\qquad
p_{\perp}=\frac{N^\prime}{2N}\!\left(\frac{m}{4\pi
r^2}-r\epsilon\right)
-\frac{\left(r^2\epsilon\right)^\prime}{2r}\,,
\end{equation}
where the function $m=m(r)$ satisfies eq. (\ref{m0}). Excluding
$N^\prime /N$ in (\ref{N}) we obtain the energy conservation law.
In order to integrate this equation it is necessary to specify the
effective matter Lagrangian or its equation-of-state.

\section{Models with integrable singularity $r=0$}\label{mode}

To give an example we consider 1st type models of eternal
black/white holes with triggered material sources with the
Lagrangian density that depends on the intrinsic curvature
$\rho\!=\!r^{-2}$ of the 2-surfaces $(r,t)\!=\!const$\,,
\mbox{$\mathcal{L}_{\mathfrak{m}}=p(\rho)$}\,. Variation of the
action with respect to the metrics (\ref{s}, \ref{R2})
yields\,\footnote{Evidently the following relation is not
universally true. For example, the functional dependance
\,\mbox{$\mathcal{L}_{\mathfrak{m}}= p(\Phi,\rho)$}\, yields
$\epsilon+p\neq 0$. For more general models see~\cite{lsm}.}
$\epsilon=-p$\,, which means that in the matter the function $N$
also remains constant enabling us to choose $N=1$ (see (\ref{N})).

Let us consider the power-law profiles of the matter density
\begin{equation}\label{m3}
\epsilon=p_0\,e\cdot\theta(1-x^{2}),\quad
e=e\!\left(x,\alpha\right)\equiv\frac{|x|^{-2\alpha}-1}{\alpha}\,,
\end{equation}
which are continuous on the borders $x\!=\!\pm 1$, where the
variable $x\equiv r/r_0$, \,the function
$\theta(y)=\int_{-\infty}\!\delta(y) dy$,\, the exponent
$\alpha=const < 3/2\,$ and the parameter $p_0$ is related to the
external mass as follows:
$$
M=4\pi r_0^3\int_{0}^{1}{\epsilon\!\left(x\right) x^2
dx}=\frac{8\pi p_{0}r_{0}^{3}}{3\left(3-2\alpha\right)}\,.
$$
The relativistic equations yield the transversal pressure
\begin{equation}
p_{\perp}\!\left(x\right)=p_0\left[1+\left(\alpha-
1\right)e\right]\cdot\theta\!\left(1-x^{2}\right),
\end{equation}
and the metric potential (\ref{rm})
\begin{equation}
\label{solution-2} \Phi\!\left(x\right)=\left\{
\begin{array}{ll}
-\displaystyle\frac{GM}{r_0}x^2\!\left(1+\frac 32 e\right),
& |x|<1 \\
\\
-\displaystyle\frac{GM}{|r|}\,, & |x|\ge 1
\end{array}.
\right.
\end{equation}

These distributions satisfy the weak energy condition while the
dominant one is broken on the matter border where the energy
density tends to zero while the transversal pressure is finite.
Note also the delta-like feature of the
distributions\,\footnote{Evaluating the limit we retain the mass
to be constant:
\[
\int_{-r_0}^{r_0}\!{\epsilon\!\left(x\right)r^2 dr}=
2\int_{-r_0}^{r_0}\!{p_{\perp}\!\left(x\right)r^2
dr}=\frac{M}{2\pi}\,.
\]
The ratio \,$\epsilon/p_\perp\stackrel{r_{0}\rightarrow
0}{\longrightarrow}2\,$ is determined by the equation-of-state
and, in general, arbitrary.} (\ref{m3})-(\ref{solution-2}) as
$r_0\rightarrow 0$\, ($\overline{\delta}(r)\equiv(4\pi
r^2)^{-1}\delta(r)$):
\begin{equation}\label{m4}
\begin{array}{ll}
\epsilon\!\left(x\right)\stackrel{r_{0}\rightarrow
0}{\longrightarrow}\,2M\overline{\delta}\!\left(r\right),\quad
p_{\perp}\!\left(x\right)\stackrel{r_{0}\rightarrow
0}{\longrightarrow}\,M\overline{\delta}\!\left(r\right),
\end{array}
\end{equation}
which proves the source of the Schwarzschild eternal black/white
hole to be localized on the hypersurface $r=0$. As mentioned
above, this matter {\it alone} cannot source the black hole, the
latter residing in the absolute past with respect to the singular
hypersurface. It is sourced by a similar matter located on the
other singular hypersurface, which lies under the point of
intersection of the horizons $r=2GM$ (see Fig. 1). In other words,
the Penrose diagram of the eternal black/white hole is an infinite
chain of elementary Penrose diagrams engaged by material regions
that source the geometry.

With $\alpha=1$, as $r\rightarrow 0$ the potential tends to a
constant, $-3GM/2r_{0}$\,, and we obtain an integrable singulary
with the divergent density $\propto r^{-2}$ and the 'hat'
transversal pressure profile, $p_{\perp}=p_0=const$ at $|r|<r_0$.
Following~\cite{MTW-3} to study the story of an extended body
falling to $r=0$, let us show that unlike the Schwarzschild
solution tidal forces in the model with $\alpha=1$ remain constant
as the body proceeds towards the singular hypersurface which
prevents mechanical disruption. Indeed, non-vanishing components
of the Riemann tensor in a locally inertial reference
frame~$(\hat{t},\hat{r},\hat{\theta},\hat{\varphi})$ are given by
the formulae:
\begin{equation}
R_{\hat{t}\hat{r}\hat{t}\hat{r}}=\Phi''\,,\qquad
R_{\hat{t}\hat{\theta}\hat{t}\hat{\theta}}=R_{\hat{t}\hat{\varphi}\hat{t}\hat{\varphi}}=\frac{\Phi'}{r}\,,
\end{equation}
\begin{equation}
R_{\hat{\theta}\hat{\varphi}\hat{\theta}\hat{\varphi}}=-\frac{2\Phi}{r^{2}}\,,\qquad
R_{\hat{r}\hat{\theta}\hat{r}\hat{\theta}}=R_{\hat{r}\hat{\varphi}\hat{r}\hat{\varphi}}=-\frac{\Phi'}{r}\,.
\end{equation}
$R_{\hat{\theta}\hat{\varphi}\hat{\theta}\hat{\varphi}}$ is the
unique component which diverges on the solution~(\ref{solution-2})
with $\alpha=1$ as $r$ tends to zero. It, however, does not enter
the equation which governs how fast two freely moving particles
separated by the spatial vector $\xi^{\hat{a}}\,$ and staying at
rest in the locally inertial reference frame accelerate one
relative to the other,
$\hat{a}=(\hat{t},\hat{\theta},\hat{\varphi})$:
$$
\frac{D^{2}\xi^{\hat{a}}}{d\hat{r}^{2}}=-R_{\hat{r}\hat{a}\hat{r}\hat{b}}\xi^{\hat{b}}\sim\frac{GM}{r_{0}^{3}}\xi^{\hat{a}}\,.
$$

Therefore, the singularity $r=0$ can be passed through and allows
us to extend the geodesics into the white-hole region.

\section{Discussion}
To summarize, we have shown that there exist geodesically complete
eternal black/white holes geometries with regular $X$-space (see
eqs. (\ref{general-metrics})-(\ref{m0}), Fig. 1) and constructed a
toy model of the geometry which depends on the intrinsic curvature
of the $Y$-space (see Sect.~6).

Based on this we argue that in the introduced class of spherically
symmetric GR solutions with 2+2 split and integrable singularity
$r=0$ in black/white holes there are solutions containing
cosmological subsystems in the $T$-regions. These include
expanding homogeneous matter flows with the spatial symmetry
$\mathbb{R}\times\mathbb{S}^{2}$. Driven by inflation they can
grow to become a Hubble flow whose parameters may be similar to
those of our Universe. In case the initial black hole originates
from collapse of a compact astrophysical object we call such
subsystem astrogenic universe. A toy model of this kind is given
in~\cite{lsm}.

Particularly, we would like to discuss the possibility of
modelling such processes. The Einstein equations imply all
physical degrees of freedom to be separated into spacetime and
material ones. In fact, effects of quantum gravity 'mix' the
degrees of freedom and the dynamical equation contain merely the
expectations $g_{\mu\nu}$ and $T_{\mu\nu}$ averaged over the
states of perturbation fields which are vacuum states in the
static $R$-region. As well known, in the quasiclassical limit the
effects of quantum gravity as well as their backreaction are
negligible. However, at large curvature it is true no more. The
emergent (non-vanishing) effective energy--momentum tensor
essentially rebuilds the original metrics extending it and
generating the 'new' space--time residing in the absolute future
relative to the 'old' world. Since the science does not have
general equations at its dispose yet while quantum corrections
following from perturbation theories contain little information
about the full picture, we simulate the 'mixing' in GR
phenomenologically. As exemplified in spherically symmetric models
we smooth out spacetime distributions (the finiteness condition
for $g_{\mu\nu}$) and consider profiles of the effective tension
$T_{\mu\nu}$ with moderate (integrable) divergence at large
curvature ($r\rightarrow 0$).

Note that the maternal matter of the collapsing stars themselves
does not constitute the astrogenic universes. It rather triggers
their generation, which dramatically differs our paradigm of
cosmic flow generation from baby-universe-like solutions and
bouncing models. Indeed, the collapsing matter is necessary to
induce the large curvature, which, in turn, triggers particle
generation in the vicinity of the singular hypersurface (see Fig.
2). If we assume that the symmetry of the Schwarzschild solution
holds inside the generated matter, the global similar symmetry
with Killing $t$-vector will be respected in the white hole as
well. Moreover, clear is the origin of the impulse that launches
the expansion -- it is the momentum accumulated in the course of
the collapse which is transferred with time outside the collapsing
object to the expanding white-hole effective matter thanks to
long-range tidal nature of gravitation.

Emphasis to be made is on extension of geodesics. The point is the
question whether it is possible to extend geodesics in a solution
has nothing to do with the existence of an astrogenic universe
(see the previous paragraph). The affirmative reply would only
mean that signals from the mother Universe can pass to the
daughter world.

The suggested hypothesis is attractive from the physical point of
view, since it allows one to relate two, perhaps, most famous GR
solutions, collapse and anticollapse (the cosmological expansion)
and resolve the initial value problem in cosmology (including the
initial symmetry) with the aid of the energy-momentum conservation
law. This paradigm of cosmogenesis requires, however, further
research -- both new models and observational tests are to be
found.

The authors thank E.V. Mikheeva and I.D. Novikov for valuable
comments. This work is supported by Ministry of Education and
Science of the Russian Federation (contracts no. P1336 of
02.09.2009 and no. 16.740.11.0460 of 13.05.2011) and by the
Russian Foundation for Basic Research (OFI 11-02-00857). V.N.S. is
grateful to the Dynasty Foundation for financial support.

\newpage

\begin{figure}[t!]
\begin{center}
{\includegraphics{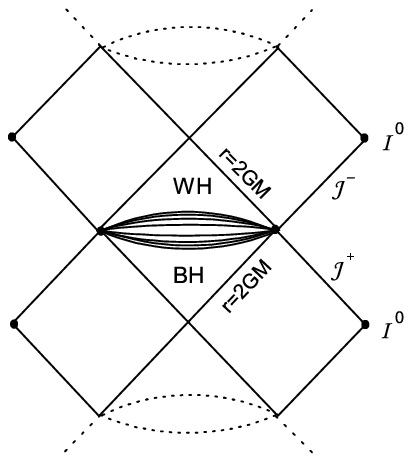}}
\end{center}
\caption{Penrose diagram of eternal black/white hole. The matter
(shaded region) separates contracting (BH) and expanding
(WH) $T$-regions}%
\label{penrose-B}
\end{figure}

\begin{figure}[t!]
\begin{center}
{\includegraphics{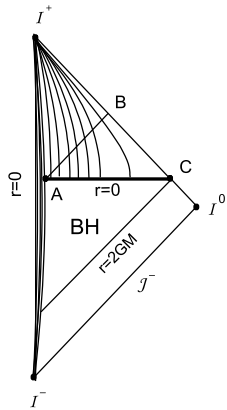}}
\end{center}
\caption{A sketch of the diagram of an astrogenic universe
($M_{WH}>>M$). $ABC$ is the region of spatially homogeneous
cosmology}%
\label{penrose-A}
\end{figure}

\end{document}